\documentclass[a4paper,superscriptaddress, 12pt,eqsecnum,floats,aps,amsmath,amssymb,nofootinbib,prd]{article}

\usepackage{geometry}
\usepackage{authblk}
\usepackage{setspace}
\usepackage{amsmath,amssymb,amsfonts,amsthm,mathrsfs}
\usepackage{graphicx}
\usepackage{enumerate} 
\usepackage{comment}
\usepackage[pdftex]{hyperref}

\def\be{\begin{equation}}
\def\ee{\end{equation}}
\def\ba{\begin{eqnarray}}
\def\ea{\end{eqnarray}}
\def\bi{\begin{itemize}}
\def\ei{\end{itemize}}

\def\bra{\langle}
\def\ket{\rangle}

\def\O{\Omega}
\def\xh{\hat{x}}
\def\scri{\mathcal{I}}
\def\R{\mathcal{R}}

\def\bo{\mathring{\beta}}
\def\vo{\mathring{V}}
\def\Uo{\mathring{U}}

\def\qo{\mathring{q}}

\def\L{\mathcal{L}}
\def\tf{\text{TF}}
\def\Hhard{H^{\rm hard}}
\def\Hsoft{H^{\rm soft}}
\def\zb{\bar{z}}

\def\G{{\cal G}}
\def\diff{{\rm Diff}}
\def\lie{{\rm Lie}}
\def\st{{\rm ST}}

\def\Go{\Gamma_0}
\def\Ocov{\Omega^{\textrm{cov}}}

\title{New symmetries for the Gravitational S-matrix}
\author[a]{Miguel Campiglia}
\author[b]{Alok Laddha}
\affil[a]{Instituto de F\'isica, Facultad de Ciencias,  Montevideo 11400, Uruguay}
\affil[b]{Chennai Mathematical Institute, Siruseri 603103, India} 
\date{}
\begin{document}
\maketitle

\let\oldthefootnote\thefootnote
\renewcommand{\thefootnote}{\fnsymbol{footnote}}
\footnotetext{E-mail: campi@fisica.edu.uy, aladdha@cmi.ac.in}
\let\thefootnote\oldthefootnote

\begin{abstract}

In \cite{us} we proposed a generalization of the BMS group ${\cal G}$ which is a semi-direct product of supertranslations and smooth diffeomorphisms of the conformal sphere. Although an extension of BMS, ${\cal G}$ is a symmetry group of asymptotically  flat space times. By taking ${\cal G}$ as a candidate symmetry group of the quantum gravity S-matrix, we argued that the Ward identities associated to the generators of $\textrm{Diff}(S^{2})$ were equivalent to the Cachazo-Strominger  subleading soft graviton theorem. Our argument however was based on a proposed definition of the $\textrm{Diff}(S^{2})$ charges which we could not derive from first principles as ${\cal G}$ does not have a well defined action on the radiative phase space of gravity. 
Here we fill this gap and provide a first principles derivation of the $\textrm{Diff}(S^{2})$ charges.
The result of this paper, in conjunction with the results of \cite{supertranslation,us} prove that the leading and subleading soft theorems 
are equivalent to the Ward identities associated to ${\cal G}$.

\end{abstract}

\maketitle

\newpage

\tableofcontents

\section{Introduction}
The space of asymptotically flat spacetimes which satisfy Einstein's equations is very rich. For example, there is an infinite dimensional symmetry group associated to this space which is known as BMS group \cite{bms,sachs}. BMS group is intrinsically tied to the  null boundary of any asymptotically flat spacetime, which in turn has a topology of $S^{2}\times{\bf R}$.  The group is a semidirect product  of an abelian group of angle-dependent translations along the null direction (referred to as `supertranslations') 
times the group of  global conformal transformations of the 2-sphere, the Lorentz group. As  asymptotically flat spacetimes have  future as well as past null infinities, the complete  group which can be associated to such spaces is a direct product $\textrm{BMS}^{+}\times\textrm{BMS}^{-}$. 
In a beautiful piece of work \cite{strominger0} Strominger introduced a remarkable notion of  ``energy conserving" diagonal subgroup $\textrm{BMS}^{0}$. It was then shown in \cite{supertranslation} that if we assume $\textrm{BMS}^{0}$ is a symmetry group of the (perturbative) quantum gravity S-matrix, then the Ward identities associated to supertranslations are in a precise sense equivalent to  Weinberg's soft graviton theorem \cite{weinberg} which relates 
 $n$-particle scattering amplitude with $(n-1)$-particle scattering amplitude
when one of the external 
particles
 is a graviton of vanishing energy.  We refer the reader to \cite{strominger0,supertranslation} for more details and for the precise definition of $\textrm{BMS}^{0}$.\\

A natural question then arises, namely if the  subleading soft graviton theorem conjectured by Strominger and proved in \cite{cs,plefka,bern} \footnote{In \cite{cs}, this theorem was  proved in the holomorphic limit. More general proofs were later given in \cite{plefka,bern}.  See \cite{gj,naculich,white} for earlier work on soft graviton amplitudes.}
 is also a manifestation of Ward identities associated to some symmetry of the perturbative S matrix. In \cite{virasoro} it was argued that the subleading soft theorem can yield Ward identities associated to `extended BMS' symmetries   \cite{bt0,bt}. The extended BMS group is a semidirect product of local (as opposed to global in the BMS case) conformal group of the two-sphere, also known as Virasoro group, and supertranslations. 
 However due to  difficulties associated to the singular nature of local conformal Killing vector fields (CKVs) on the conformal sphere, it was not clear how to obtain the subleading soft theorem from Ward identities. \\ 

In \cite{us}, motivated by the precise equivalence between supertranslation Ward Identities and Weinberg's soft graviton theorem, we argued that the Cachazo-Strominger (CS) theorem\footnote{Based on earlier papers, we refer to the subleading soft theorem as Cachazo-Strominger theorem.} was in fact equivalent to Ward identities associated not to the Virasoro group but to the group of sphere diffeomorphisms $\diff(S^{2})$ which belongs not to the `extended BMS' group but to 
what we called `generalized BMS' group $\G$. This group has the same structure as the (extended) BMS group,  but instead of (local) CKVs of the conformal sphere one allows for arbitrary smooth sphere vector fields. That is,  $\G$ is a  semidirect product of diffeomorphisms of the conformal sphere and supertranslations. 
Based on earlier literature on radiative phase space and asymptotic symmetries \cite{aaprl,as,aajmp,aaoxf,aabook,bt0,bt} it was argued in \cite{us} that $\G$ is a  symmetry of Einstein's equations (with zero cosmological constant) if one allows for arbitrary metrics on the conformal sphere. We then showed that if the charges\footnote{Here `charge' refers to what is called `flux' in the radiative phase space literature, i.e. it involves a three dimensional integral over null infinity.} of the $\diff(S^2)$ generators 
were exactly equal to the charges of Virasoro generators given in \cite{virasoro} then the Cachazo-Strominger theorem was equivalent to the Ward identities associated to $\diff(S^2)$ \\

A key question left unanswered in \cite{us} was whether the proposed charges could be derived from canonical methods.  The difficulty stem from the fact that the action of $\G$ does not preserve Ashtekar's radiative phase space (see \cite{aa2014} for a recent review).  More in detail, the radiative phase space $\Gamma^{q}$ depends an a  given choice of sphere 2-metric $q_{AB}$ or `frame'\footnote{We are deviating from the standard radiative phase space terminology in which `frame' denotes a conformal class of  $[(q_{ab},n^a)]$ of metric \emph{and} null normal \cite{as}.}  at null infinity. 
In contrast to the BMS group, $\G$ does not preserve $\Gamma^{q}$ since the $\diff(S^2)$ factor does not preserve the given frame $q_{AB}$.   Thus, the strategy that had successfully lead to BMS charges \cite{as} could not be applied here. \\

It is then natural to attempt to work in the space $\Gamma \sim \cup_{\{ q \}} \Gamma^{q}$ off \emph{all} radiative phase spaces on which  $\G$ acts in a well defined manner.   However we were so far  lacking a symplectic structure on $\Gamma$.   It is here that we turn to covariant phase space methods \cite{abr,lw}.

It is well known \cite{am} that the symplectic structure on $\Gamma^{q}$ corresponds to the GR covariant phase space symplectic  structure $\Ocov$ evaluated at null infinity. Here we will show that $\Ocov$ naturally defines a  symplectic structure on (a suitable subspace of) $\Gamma$. By realizing $\Gamma^{\qo}$ as a  symplectic subspace of $\Gamma$, we will be able to derive the $\G$-charges that were postulated in \cite{us}.  
\\

The outline of the paper is as follows. Section \ref{sec2}  provides the background material for our discussion. In \ref{sec2.1} we describe the class of spacetimes under consideration following closely reference \cite{bt}. In \ref{sec2.2} we  review the definition of generalized BMS group $\G$ as a symmetry group of such spacetimes. In \ref{sec2.3} we recall the definition of radiative phase space associated to an arbitrary `frame' an introduce the total space $\Gamma$ of all such radiative phase spaces. We also introduce certain subspaces with stronger fall-offs in $u$  that play a crucial role in the later discussion.

Section \ref{sec3} is the main part of the paper.  \ref{sec3.1} describes the general idea behind our computation. In  \ref{sec3.2} we show that the covariant phase space symplectic structure induces a symplectic structure on (suitable subspace of)  $\Gamma$.  In section \ref{sec3.3} we use this symplectic structure to derive the charges associated to the generators of $\diff(S^2)$. This represents the main result of the paper.  In  \ref{sec3.4} we summarize the analogue  results at  past null infinity (our detailed calculations  take place in future null infinity). Finally in  \ref{sec3.5} we give a brief summary of the results presented in \cite{us} on the equivalence between the $\diff(S^2)$ Ward identities and the CS theorem.

In section \ref{sec4} we argue that subleading soft gravitons can be thought of as Goldstone modes of a spontaneous symmetry breaking $\G \to \textrm{BMS}$, in complete parallel to how leading soft gravitons are thought  as  Goldstone modes of a spontaneous symmetry  breaking  from supertranslations to translations  \cite{supertranslation}.

We end with the conclusions in section \ref{sec5}.\\

\section{Preliminaries} \label{sec2}
\subsection{Spacetimes under consideration} \label{sec2.1}
As in \cite{strominger0,supertranslation,virasoro} we are interested in spacetimes that are asymptotically flat at both future and past null infinity.  For concreteness we focus on the description from future null infinity; similar considerations apply to the description from past null infinity.  We follow closely Reference \cite{bt}. 

In Bondi coordinates $(u,r,x^A)$ the 4-metric is parameterized as  
\be
ds^2 = (V/r) e^{2 \beta}  du^2 - 2 e^{2\beta} du dr + g_{AB}(dx^A-U^A du)(dx^B - U^B du) , \label{4metric}
\ee
with $\beta,V/r,U^A$ and $g_{AB}$ satisfying the $r\to \infty$ fall-offs
\be
\beta = r^{-2}\bo + O(r^{-3}) , \quad V/r= \vo + r^{-1} 2 M +O(r^{-2}), \quad U^A= r^{-2} \Uo^A +O(r^{-3}) \label{falloffs}
\ee
\be
g_{AB}= r^{2}q_{AB} + r C_{AB} + \frac{1}{4} q_{AB} C^2 +O(r^{-1}). \label{gAB}
\ee
Here $C^2 \equiv C_{AB} C^{AB}$ (sphere indices are raised and lowered with $q_{AB}$) and the coefficients of the $1/r$ expansion in (\ref{falloffs}) and (\ref{gAB}) are functions of $u$ and $x^A$ except for $q_{AB}$ which,  in contrast to \cite{bt}, we assume to be $u$-independent.  There is an additional  gauge fixing condition 
\be
\det(g_{AB})= r^4\det(q_{AB})
\ee
which in particular implies that $q^{AB}C_{AB}=0$ and that the trace part of the $O(1)$ term of $g_{AB}$ has the form given in (\ref{gAB}). We take the trace-free $O(1)$ part of $g_{AB}$ to be zero as in the original treatment by Sachs (see discussion following Eq. (4.38) in \cite{bt}). 

An important difference with the treatment of \cite{bt} is that we do not demand $q_{AB}$ to be proportional to the unit round metric $\qo_{AB}$. So far it can be \emph{any} sphere metric. One can verify that Einstein equations still imply the relations given in  Equations (4.42), (4.36) and (4.37) of \cite{bt}:
\be
\vo = - \frac{1}{2} \R , \quad \bo= - \frac{1}{32} C^2 , \quad U^A= - \frac{1}{2} D_B C^{AB},
\ee
where $\R$ and $D_A$ are respectively the scalar curvature and covariant derivative of $q_{AB}$. Finally, as it will become clear in the next subsection,  the natural space of metrics to consider from the point of view of the generalized BMS group $\G$ is one where the area form of $q_{AB}$ is fixed: $\sqrt{q}= \sqrt{\qo}$. 

To summarize, we will be interested in spacetime metrics of the form (\ref{4metric}) parametrized by `free data'  $(q_{AB},C_{AB})$ \footnote{This actually is not the totality of free data since there are additional $u$-independent sphere functions that arise as integration `constants' \cite{bt}. These however  play no role in our analysis.}  
  satisfying 
\be
 \partial_u q_{AB}=0, \quad  \sqrt{q}= \sqrt{\qo}, \quad  q^{AB} C_{AB} =0. \label{fdspace}
\ee
In section \ref{fallC} we describe  conditions on $C_{AB}$ as $u \to \pm \infty$.

\subsection{Definition of $\G$} \label{sec2.2}
As described in \cite{us}, from a spacetime perspective $\G$ can be characterized as the group of  diffeomorphisms generated by (non-trivial at null infinity) vector fields $\xi^a$ preserving the form of the metric (\ref{4metric}) and such that they are asymptotically divergence-free (instead of asymptotically Killing as in the BMS case). Such vector fields are parametrized by a sphere function $f(\xh)$ (supertranslation) and sphere vector field $V^A(\xh)$ according to \cite{bt,us}:
\be
\xi^a_f = f \partial_u + \ldots , \quad  \;   \xi_V^{a} =V^{A} \partial_A + u \alpha  \partial_u  - r \alpha \partial_r + \ldots \label{xiVf}
\ee
where $\alpha = (D_C V^C)/2$ and the dots indicate subleading term in the $1/r$ expansion that  depend on $f$ and $V$ and in the 4-metric `free data'.  The relations defining the algebra $\lie(\G)$ are obtained by computing the leading terms of the Lie brackets of the vector fields (\ref{xiVf}). One finds:
\be
[\xi_{f_1},\xi_{f_2}] =0, \quad  [\xi_{V_1}, \xi_{V_2}] = \xi_{[V_1,V_2]}, \quad [\xi_{V}, \xi_{f}] = \xi_{\L_V f- \alpha f} .\label{algebra}
\ee
Thus as in the (extended) BMS case, $\lie(\G)$ has a semidirect sum algebra structure, where supertranslations form an  abelian ideal $\lie(\st)$ and  $\lie(\G)/\lie(\st)$  is the algebra of sphere vector fields. Similarly to the BMS case, one can also characterize the group $\G$ as diffeomorphisms of an abstract $\scri$ preserving certain structure (see section 4.1 of \cite{us}).

By computing the Lie derivative of the metric (\ref{4metric}) along the vector fields (\ref{xiVf}) one obtains the following action of $\lie(\G)$ on the free data \cite{bt,us}:
\be
\delta_f q_{AB} = 0 , \quad \delta_f C_{AB} = f \dot{C}_{AB} - 2 (D_A D_B f)^{\tf}
\ee
\be
\delta_V q_{AB}  =  \L_V q_{AB}-2 \alpha q_{AB} , \quad \delta_V C_{AB} =  \L_V C_{AB} - \alpha C_{AB} + \alpha u \dot{C}_{AB}  - 2 u (D_A D_B \alpha)^{\tf} , \label{deltaV}
\ee
where `$\tf$' denotes trace-free part with respect to $q_{AB}$. In appendix \ref{appclosure} we verify this action indeed reproduces the algebra (\ref{algebra}):
\be
[\delta_{f_1},\delta_{f_2}] =0, \quad  [\delta_{V_1}, \delta_{V_2}] = - \delta_{[V_1,V_2]}, \quad [\delta_{V}, \delta_{f}] =-  \delta_{\L_V f- \alpha f}. \label{algebrafd}
\ee
Here  $\delta_f$ and $\delta_V$ are understood as vector fields on the space of free data $\{ (q_{AB},C_{AB}) \}$ satisfying (\ref{fdspace})  (strictly speaking  $-\delta_f$ and $-\delta_V$ are the vector fields that provide the representation of the algebra (\ref{algebra})).  \\

Similar analysis on past null infinity yields a generalized BMS group associated to $\scri^-$. 
Thus the total group acting on the spacetimes we are interested in is $\G^+ \times \G^-$, ($\G^+$ is what we have been calling  $\G$). The proposed symmetry group of the gravitational S matrix is the `diagonal' subgroup $\G^0 \subset \G^+ \times \G^-$ defined in analogy to Strominger's $\textrm{BMS}^0$ \cite{us}.

\subsection{Radiative phase spaces}  \label{fallC} \label{sec2.3}

We first recall the asymptotic conditions on $C_{AB}$ that ensure well-definedness of the radiative phase spaces \cite{as}.  The radiative phase space $\Gamma^{q}$ associated to a sphere metric $q_{AB}$ is given by tensors $C_{AB}$ on $\scri$ satisfying:
\begin{equation}
\Gamma^{q}\ :=\ \{ C_{AB} : \;\ q^{AB} C_{AB} =0, \quad  C_{AB}(u,\xh) = u (\rho_{AB})^{\tf}+ C^{\pm}_{AB}(\xh) + O(u^{-\epsilon}) \}, \label{defGq}
\end{equation}
where $\epsilon>0$. Here $\rho_{AB}$ is a fixed tensor that depends on $q_{AB}$;  its definition is reviewed in section \ref{ssb1}.  The radiative phase space traditionally used in the literature is the one associated to the unit round metric $\qo_{AB}$ on which $(\rho_{AB})^\tf =  0$.  \\

We  define $\Gamma$ as the union of all $\Gamma^q$ spaces with given area element $\sqrt{q}=\sqrt{\qo}$:
\be
\Gamma := \bigcup_{\sqrt{q}=\sqrt{\qo}} \Gamma^q. \label{defGamma}
\ee
The properties of $\rho_{AB}$ in (\ref{defGq}) ensure that the  action (\ref{deltaV}) preserves the form of the  linear in $u$  term in (\ref{defGq}),  so that indeed  $\G$ has a well defined action on $\Gamma$. The precise mechanism by which this occurs is described in section \ref{ssb1}. \\

As in  \cite{us}, due to infrared issues, the charges associated to (extended and) generalized BMS group will be defined on the following subspace of $\Gamma^{\qo}$:
\begin{equation}
\begin{array}{lll}
\Gamma^{\qo}_0 := \{  C_{AB}  \in \Gamma^{\qo} : C_{AB}(u,\xh)=  O(u^{-1-\epsilon}) \quad \text{as } \; u \to \pm \infty \}. \label{defGqoo}
\end{array}
\end{equation}


We similarly define a subspace  $\Gamma$ on which  the covariant phase space symplectic structure will turn out to be well-defined:
\be
\Gamma_0:=\{ (q_{AB},C_{AB}) : \; \partial_u q_{AB}=0, \;  \sqrt{q}= \sqrt{\qo}, \; q^{AB} C_{AB} =0, \;  C_{AB}(u,\xh) =   O(u^{-1-\epsilon}) \} . \label{defGo}
\ee

The spaces (\ref{defGqoo}) and (\ref{defGo})  will play an essential role in ensuring integrals in $u$ are finite. We would like to emphasize however that this way of avoiding IR divergences is not  entirely satisfactory: $\Gamma_0$ is not preserved by $\G$ and $\Gamma^{\qo}_0$ is not preserved by supertranslations\footnote{The analogue of the space $\Gamma^{\qo}_0$ that was used in \cite{us} actually allows for a $u$-independent term and hence is invariant under supertranslations. The stronger condition (\ref{defGqoo}) is used here in order to allow for certain integration by parts in $u$}.  There may be better ways of dealing with these IR issues, for instance by introducing appropiate counterterms  (see footnote \ref{countertermfn}).  We hope to return to this point  in future investigations.


\section{Main section} \label{sec3}
\subsection{General idea} \label{sec3.1}
In this section we show that starting from the 
covariant phase space derived from the Einstein Hilbert action, one can obtain a phase space 
at null infinity which is coordinatized not only by radiative degrees of freedom $C_{AB}$ but also by the 2-metric $q_{AB}$ on the conformal sphere. 
It turns out that the radiative phase space 
is a symplectic subspace of this larger space. 
This will allow us to compute the corresponding charge, which is well defined on a (suitable) subspace of $\Gamma^{\qo}$, where $\qo_{AB}$ is the unit round-metric on the 2 sphere. This is the main result of the paper, which combined with the result of \cite{us} show that Ward identities associated to $\G$ are equivalent to the CS soft theorem.\\


The main idea can be summarized as follows. Let 
\be
\O_{t,g}(\delta ,\delta'  ) := \int_{\Sigma_t} dS_a \omega_g^a(\delta,\delta'), \label{Ot}
\ee
 be the standard covariant phase space symplectic form \cite{abr,lw} evaluated on a $t := r+ u=$constant slice $\Sigma_t$.   If we characterize 4-metrics $g_{ab}$ by the free data  $(q_{AB},C_{AB})$, the $t \to \infty$ limit of (\ref{Ot}) could correspond to a symplectic product defined at $\scri^+$. However, one needs to impose conditions on the given fields $(q_{AB},C_{AB}, \delta, \delta')$ in order for this limit to be well defined. For instance, for variations $\delta,\delta'$ such that $\delta q_{AB}=\delta' q_{AB}=0$ and such that $\delta C_{AB}, \delta' C_{AB}$ satisfies appropriate fall-offs in $u$, this procedures reproduces  Ashtekar's radiative phase space  symplectic structure \cite{am}.   In section \ref{sstr} we show that if  $C_{AB}$ and its variation are taken to be $O(u^{-1-\epsilon})$ and $q_{AB}$ is allowed to vary, then (\ref{Ot}) also has a well defined $t \to \infty$ limit. In other words, the covariant phase space symplectic form induces a well defined symplectic form on the space $\Gamma_0$ defined in Eq. (\ref{defGo}).

This is not quite yet what we need, since $\delta_V \notin T \Gamma_0$ due the linear in $u$ term in (\ref{deltaV}).   We now explain the computation we are really interested in. 
Given the phase space $\Gamma^{\qo}_{0}$ and given a symmetry generator $V \in \lie(\G)$, we would like to compute its associated charge $H_{V}$ as a function on $\Gamma^{\qo}_{0}$. Now in general, given a Hamiltonian Vector field on any phase space, we first compute the differential of the corresponding Hamiltonian function, which upon integration yields the corresponding Hamiltonian. This implies that in our case, given the action of $\delta_{V}$ 
what we would really like to compute is 
\begin{equation}
\O(\delta_{V},\delta)\ =:\ \delta H_{V}  \label{defHV}
\end{equation}
where the variation $\delta$ must be along $\Gamma^{\qo}_{0}$. 
As we will see,  the symplectic product needed in  (\ref{defHV}) is still well defined. In other words Eq. (\ref{Ot}) has also a well defined $t \to \infty$ limit when $\delta =\delta_V$ and $\delta' \in T \Gamma^{\qo}_{0}$.


\subsection{Symplectic structure on $\Gamma_0$} \label{sstr} \label{sec3.2}
In this section we evaluate  the $t \to \infty$ limit of (\ref{Ot}) on $\Go$. The computation simplifies  by working with the symplectic potential
\be
\Theta_{t,(q,C)}(\delta) := \int_{\Sigma_t} dS_a \theta^a,
\ee
\be
\theta^a :=\frac{1}{2} \sqrt{g} \big( g^{bc} \delta \Gamma^a_{bc} - g^{a b} \delta \Gamma^c_{c b} \big).
\ee
Recall $t=r+u$ so that the relevant component will be $\theta^t = \theta^r+\theta^u$.   The limit is taken $t \to \infty$ with $u$ constant.  The variable $r$ will be understood as given by $r=t-u$. \\

Before proceeding with the details, we summarize certain salient aspects of the computation which critically use the fall-off conditions on $C_{AB}$.\\
\noindent{\bf (a)} We will see that in the limit $t \to \infty$,  conditions $C_{AB}=O(u^{-1-\epsilon})$ and $\delta \in T\Gamma_0$ will ensure finiteness of the integrals as well as cancellation of various boundary terms.\footnote{For slower fall-offs of the type which define $\Gamma$ (Eqns.  (\ref{defGq}), (\ref{defGamma})) there may be a possibility of obtaining a finite symplectic structure by supplementing  the action with counterterms at $i^{0}, i^{\pm}$.
We have not pursued this direction here as it is not needed for our analysis. \label{countertermfn}}  
We will discard total variation terms, since they do not contribute to the symplectic form.

\noindent{\bf (b)} We will find that $\theta^t$ has an $1/r$ expansion of the form:
\be
\theta^t = r \theta^t_1 + \theta^t_0 + O(r^{-1}),
\ee
which  gives the following  $1/t$ expansion:
\be
\theta^t = t \theta^t_1 + (\theta^t_0 - u \theta^t_1) + O(t^{-1}).
\ee
The would-be divergent term $t \theta^t_1$ will turn out to integrate to zero on the space of $C$'s we are restricting attention to.\\

We now proceed with the details of the computation.

The form of the metric (\ref{4metric}) implies the non-zero term appearing in $\theta^t$ are:
\begin{multline}
2 \theta^t = \sqrt{q} e^{2 \beta} r^{2} \big( 2 g^{ur} \delta \Gamma^r_{ur} + g^{rr}\delta \Gamma^r_{ur}+ 2 g^{Ar}\delta \Gamma^r_{A r} + g^{AB}(\delta \Gamma^r_{AB}+ \delta \Gamma^u_{AB})  \\ - g^{ur}(\delta \Gamma^c_{cu}+ \delta \Gamma^c_{c r}) -g^{Ar} \delta \Gamma^c_{cA} \big). \label{thetat}
\end{multline}
There are 6 terms in (\ref{thetat}), lets call them (1)...(6). Setting to zero terms that are $O(r^{-1})$ we get:
\be
(1)= -2 \sqrt{q} \delta M , \quad (2)= (3)=0, \quad  (5)= 2 \sqrt{q}\delta \dot{\bo}, \quad (6) =  \Uo^A \partial_A \delta(\sqrt{q})
\ee
For the  space we are interested where $\sqrt{q}$ is fixed, the term (6) vanishes and the terms (1) and  (5)  are total variations which do not contribute to the symplectic structure.  The only nonzero term is the fourth one in (\ref{thetat}) so we rewrite $\theta^t$ as:

\be
 \theta^t =\frac{1}{2} \sqrt{q} e^{2 \beta} r^{2}  g^{AB}(\delta \Gamma^r_{AB}+ \delta \Gamma^u_{AB}) . \label{thetat2}
\ee
The relevant Christoffel symbols are (see \cite{bt}):
\ba
\Gamma_{AB}^r & = &  D_{(A} \Uo_{B)} + \frac{r}{2} \dot{C}_{AB} +\frac{1}{8}q_{AB} \partial_u (C^2)+ \vo r q_{AB} + \frac{1}{2} \vo C_{AB}+ 2 M q_{AB} +O(r^{-1})\label{GammarAB} \\
\Gamma^{u}_{AB} &= & -\frac{1}{2} g^{ur} \partial_r g_{AB} = r q_{AB} + \frac{1}{2} C_{AB} +O(r^{-1}).
\ea
We now evaluate the first term in (\ref{thetat2}):
\ba
r^{2}e^{2\beta}g^{AB} \delta \Gamma^{r}_{AB} =  q^{AB}\delta \Gamma^{r}_{AB} -r^{-1} C^{AB} \delta \Gamma^{r}_{AB} +O(r^{-1}). \label{thetar}
\ea
In evaluating (\ref{thetar}) we will discard terms of the form $q^{AB} \delta (f q_{AB})$ for any quantity $f$ since they give a total variation by the identity
\be
\sqrt{q}q^{AB} \delta (f q_{AB}) =  \delta (\sqrt{q} f)
\ee
that follows from  $q^{AB}q_{AB}=2$ and $2 \delta \sqrt{q}= \sqrt{q} q^{AB} \delta q_{AB}$.
Substituting (\ref{GammarAB}) in (\ref{thetar}) we get
\be
r^{2}e^{2\beta}g^{AB} \delta \Gamma^{r}_{AB} = q^{AB} \delta( D_A \Uo_B) - \frac{1}{2} C^{AB}\delta \dot{C}_{AB} -\frac{1}{2} \vo C^{AB}\delta q_{AB} + \frac{r}{2}q^{AB} \delta \dot{C}_{AB} + \delta( ) +O(r^{-1}) \label{thetar2}
\ee
where $\delta()$ indicates a total variation term. 
When we substitute $r=t-u$, the linear in $r$ term in (\ref{thetar2}) gives a potential diverging linear in $t$ term and a finite term 
\be
- \frac{u}{2}q^{AB} \delta \dot{C}_{AB}  =  \frac{1}{2} C^{AB} \delta q_{AB}  - \partial_u(\frac{u}{2}\delta C_{AB})q^{AB} + \delta() \label{uqC}
\ee
which we rewrote up to total derivative in $u$  and a total variation. \emph{Now, the condition $C_{AB}=O(u^{-1-\epsilon})$ implies  the total derivative in (\ref{uqC}) as well as the potential diverging term $\frac{t}{2}q^{AB} \delta \dot{C}_{AB}$ give a vanishing contribution upon integration.} 

The second term in (\ref{thetat2}) gives:
\ba
r^{2}e^{2\beta}g^{AB} \delta \Gamma^{u}_{AB} & = & q^{AB}\delta \Gamma^{u}_{AB} -r^{-1} C^{AB} \delta \Gamma^{u}_{AB} +O(r^{-1}) \\
& = & -\frac{1}{2} C^{AB}\delta q_{AB} + \delta()+O(r^{-1})
\ea
Note that this term cancels the term  in (\ref{uqC}). Collecting all terms and writing for later convenience 
\be
 q^{AB} \delta( D_A \Uo_B) =  D^A \Uo^B \delta q_{AB} + \delta() 
\ee
we obtain the following  expressions for the symplectic potential $\Theta(\delta):= \lim_{t \to \infty} \Theta_{t}(\delta)$ at $\scri^+$:
\be
\Theta(\delta)= \frac{1}{4}\int_{\scri} du \sqrt{q} \left(- C^{AB} \delta \dot{C}_{AB}  + \big[2 D^{A} \Uo^B - \vo C^{AB} \big]\delta q_{AB} \right).
\ee
The corresponding symplectic form at $\scri^+$ is then:
\be
\O(\delta,\delta') = \frac{1}{4}\int_{\scri} du \sqrt{q} \left(\delta C^{AB} \delta' \dot{C}_{AB}  - \delta (2 D^{A} \Uo^B - \vo C^{AB} ) \delta' q_{AB} \right)   - \delta \leftrightarrow \delta'. \label{sf}
\ee
We have thus obtained a symplectic form on the space $\Gamma_0$ defined in Eq. (\ref{defGo}). Clearly, the radiative phase space  $\Gamma^{\qo}_0$ is  symplectic subspace of $\Gamma_0$.\\

We conclude with  the observation that (\ref{sf}) can actually be used for the evaluation of the symplectic product between 
$\delta_V \in T \Gamma$ and $\delta_0 \in T \Gamma^{\qo}_0$. 

By introducing a second variation in all steps above, one can verify that: 
\be
\lim_{t \to \infty} \O_t(\delta_V,\delta_0) = \O(\delta_V,\delta_0),
\ee
 with $\O$ given in (\ref{sf}). Indeed, there are only two potentially problematic terms in the computation that are described after Eq. (\ref{uqC}). Their contribution to the  density $\omega^t(\delta_V,\delta_0)$ is:
\be
-\frac{t}{2}\delta_V q^{AB} \delta_0 \dot{C}_{AB} + \partial_u(\frac{u}{2}\delta_0 C_{AB}) \delta_V q^{AB},\label{wdiv}
\ee
where we used that $\delta_0 q_{AB}=0$. The condition $\delta_0 C_{AB}= O(u^{-1-\epsilon})$ implies that both terms in (\ref{wdiv}) integrate to zero. In summary,  the symplectic product between $\delta_V$ and $\delta_0 \in T \Gamma^{\qo}_0$ is well defined and given by evaluation on the form (\ref{sf}). This  evaluation is used in the next section to obtain the charge $H_V$.

\subsection{$\diff(S^{2})$ charges} \label{sec3.3}
We now apply the above results to find the charge $H_V$ satisfying
\be
\delta H_V = \O(\delta_V, \delta) ,\label{hvfV}
\ee
for  $\delta_V$  given in Eq. (\ref{deltaV}) and for  $\delta \in T \Gamma^{\qo}_0$, i.e.  $\delta C_{AB}=O(u^{-1-\epsilon})$ and $\delta q_{AB}=0$. Since we already have a candidate for $H_V$, namely the one postulated in \cite{us}, we will just verify that such $H_V$ indeed satisfies (\ref{hvfV}). 


$H_V$  is a sum of a `hard'  quadratic in $C_{AB}$ term and a `soft' linear in $C_{AB}$ term \cite{us}:
\be
H_V = \Hhard_V + \Hsoft_V ,\label{Hhs}
\ee
\ba
\Hhard_V & := & \frac{1}{4} \int du  \sqrt{q} \,\dot{C}^{AB}(\L_V C_{AB} - \alpha C_{AB} + \alpha u \dot{C}_{AB}) \label{Hhard}\\
\Hsoft_V & := & \frac{1}{2} \int du  \sqrt{q}  \, C^{AB}  s_{AB}, \label{Hsoft}
\ea
with $s_{AB}$ a symmetric trace-free tensor such that its components in  $(z,\zb)$ coordinates are given by:
\be
s_{zz} := D^3_z V^z \label{szz},
\ee
and corresponding complex conjugated expression (the trace-free condition sets $s_{z \zb}=0$).
We now verify that $H_V$ satisfies (\ref{hvfV}). \\

For the RHS of (\ref{hvfV}) we have
\be
\O(\delta_V,\delta) =  \frac{1}{4}\int du \sqrt{q} \left(\delta_V C^{AB} \delta \dot{C}_{AB} - \delta C^{AB} \partial_u (\delta_V C_{AB})  +  (2 D^{A} \delta \Uo^B + \delta C^{AB} ) \delta_V q_{AB} \right) , \label{OddV}
\ee
where we used that $\delta q_{AB}=0$ and  $\vo = -\R/2=-1$ since we are at $q_{AB}=\qo_{AB}$. 
Using  (\ref{deltaV})  and the corresponding transformations:
\be
\delta_V C^{AB} =  \L_V C^{AB} + 4 \alpha C^{AB} + \alpha u \dot{C}^{AB}  - 2 u (D^A D^B \alpha)^{\tf} \label{delVCcont}
\ee
\be
\delta_V q^{AB}  =  \L_V q^{AB}+2 \alpha q_{AB} \label{delVqinv},
\ee
one verifies that the `hard' terms in (\ref{OddV}) combine to give $\delta \Hhard_V$. By integration by parts  one can bring all `soft' terms in a form that is proportional to $\delta C^{AB}$. The end result is:
\be
\O(\delta_V,\delta) = \delta \Hhard_V +\frac{1}{2} \int du  \sqrt{q}  \, \delta C^{AB}  s'_{AB}
\ee
where
\be
s'_{AB} :=  ( 2 D_A D_B \alpha - \frac{1}{2} D_{(A} D^M \delta_V q_{B) M} + D_{(A} V_{B)})^\tf. \label{sp}
\ee
We finally show that $s'_{AB}=s_{AB}$ from which  (\ref{hvfV}) follows.

From (\ref{deltaV}) and using the identity $D^M D_B X_M = X_B + 2 D_B \alpha$ one finds
\be
D^M \delta_V q_{B M} =  \Delta V_B + V_B.
\ee
Using this we can rewrite (\ref{sp}) as
\be
s'_{AB} = (D_{(A} s'_{B)})^{\tf} \label{Dsp}
\ee
with
\be
s'_A :=  D_A D_M V^M - \frac{1}{2} D_M D^M V_A + \frac{1}{2} V_A. \label{spA}
\ee
Finally, writing (\ref{spA}) in $(z,\zb)$ coordinates and using $\qo^{z \zb} [D_{\zb},D_z] V_z = V_z$ one finds
\be
s'_z = D^2_z V^z .
\ee
Going back to (\ref{Dsp}) we conclude that $s'_{AB}=s_{AB}$ as desired. \\

\noindent \emph{Comment:} In order to highlight the  role played by the `extra' terms in the symplectic structure (\ref{sf}),  it is interesting to repeat the computation by writing the symplectic structure as $\O(\delta,\delta') = \frac{1}{4}\int  \left( \delta C^{AB} \delta' \dot{C}_{AB} + \delta(2 a \,D^{(A} \Uo^{B)} + b \,  C^{AB} ) \delta' q_{AB} \right) - \delta \leftrightarrow \delta'$  and setting the correct values $a=-1,b=\vo=-1$ at the end of the computation (for the $\delta$ considered here $\delta \vo=0$ and so  $\vo$  can be treated as a constant).   Doing so one obtains: $s'_z = D^2_z V^z + (1+ a) D_z D_{\zb} V^z + (a-b) V_z$.  \\

\subsection{Past null infinity} \label{sec3.4}

A similar analysis to the one given in the previous two subsections goes through for past null infinity. The form of the metric in that case  can be obtained by doing the substitution $v = -u$ in (\ref{4metric}). The relevant component of the symplectic potential density is now $\theta^t= -(\theta^r - \theta^v)$. Thus, the symplectic structure at past null infinity can be obtained by the replacement $u \to -v$ (up to an overall sign). The result is:
\be
\O^-(\delta,\delta') = \frac{1}{4}\int_{\scri} dv \sqrt{q} \left(\delta C^{- AB} \delta' \dot{C}^-_{AB}  + \delta (2 D^{A} \Uo^{- B} - \vo^- C^{- AB} ) \delta' q^-_{AB} \right)   - \delta \leftrightarrow \delta'. \label{sfm}
\ee
On the other hand, the transformation rule for $C^-_{AB}$ is the same as for $C^+_{AB}$ except that the soft factor  comes with opposite sign:
\be
\delta_V C^-_{AB} =  \L_V C^-_{AB} - \alpha C^-_{AB} + \alpha v \dot{C}^-_{AB}  + 2 v (D_A D_B \alpha)^{\tf}.
\ee
The corresponding charge $H^-_V$ has thus the same form as (\ref{Hhs}), (\ref{Hhard}), (\ref{Hsoft}), with an opposite sign in the soft term:
\be
s^-_{zz}= -D^3_z V^z .
\ee

\subsection{$\diff(S^2)$ Ward identities and CS soft theorem} \label{sec3.5}
We sketch here how the new symmetry relates to CS soft theorem. We refer to \cite{virasoro,us} for further details. 

Given a vector field $V^A$ and corresponding charges $H^\pm_V$ at future and past null infinity, the proposed Ward identities arise from assuming the S matrix satisfies:
\be
H^+_V S = S H^-_V  ,\label{hssh}
\ee
or equivalently:
\be
H^{\textrm{soft}  +}_V S -  S H^{\textrm{soft}  -}_V = - H^{\textrm{hard}  +}_V S +  S H^{\textrm{hard}  -}_V. \label{wardid}
\ee

When one takes the matrix element of (\ref{wardid}) between a  $n^+$ particle state $\bra {\rm out} |$ and a $n^-$ particle state $| {\rm in} \ket$, the RHS of (\ref{wardid}) becomes an  operator on the  scattering amplitude $\bra {\rm out} | S | {\rm in} \ket$ that consists of a sum of differential operators acting on the individual particle labels (momentum and helicity). On the other hand, the LHS of (\ref{wardid}) can be realized as creation operators of gravitons with vanishing energy, where the helicity and  smeared momentum direction is determined by $V^A$.

The choice
\be
V^A(z,\zb) = K^A_{(z_s, \zb_s)}(z,\zb) := (\zb-\zb_s)^{-1}(z-z_s)^2\partial_{z} \label{VK}
\ee
gives, in the first term of the LHS of (\ref{wardid}), the insertion of a negative helicity outgoing soft graviton with momentum pointing in the direction determined by $(z_s,\zb_s)$.  By crossing symmetry the second term in the LHS of (\ref{wardid}) can be shown to be equal to the first one. Now, the differential operators arising on the RHS of (\ref{wardid}) for the choice (\ref{VK}) reproduce those of the CS theorem. In short, for $V^A=K^A_{(z_s, \zb_s)}$, Eq. (\ref{wardid}) reproduces CS soft theorem for a negative helicity graviton (the positive helicity case  is obtained by choosing the complex conjugated vector,  $V^A= \bar{K}^A_{(z_s,\zb_s)}$).

Conversely, the Ward identities associated to the vector fields  (\ref{VK}) and its complex conjugate (which we just argued are equivalent to CS theorem), can be shown to  imply the  Ward identity (\ref{wardid}) for \emph{any} vector field $V^A$. Essentially the vectors $K^A_{(z_s, \zb_s)}(z,\zb)$ have the role of elementary kernels, and by appropriate smearing in the $(z_s,\zb_s)$ variables one can reproduce any desired vector field.

\section{Goldstone modes of ${\cal G}$} \label{ssb} \label{sec4}
In the case of supertranslation symmetry, it was argued in \cite{supertranslation} that 
as supertranslations map an asymptotic configuration $C_{AB}$ with zero news $N_{AB}\ =\ 0$ to a distinct configuration with zero news (by creating a soft graviton),  the  choice of a particular vacuum implies a spontaneous breaking of supertranslation symmetry with soft gravitons playing the role of Goldstone modes. \\
In this section we argue that one can interpret the subleading soft gravitons in a similar manner and that they can be thought of as Goldstone modes associated to spontaneous breaking of ${\cal G}$ to $\textrm{BMS}$. At first sight this statement looks obviously wrong as for a given choice of the sphere metric, such subleading changes in $C_{AB}$ are not gapless. This can be seen as follows. Given  $(C_{AB}, q_{CD})\ \in \Gamma$,   a vector field $V\ \in \textrm{Lie}({\cal G})$ maps it to
\begin{equation}
\begin{array}{lll}
\delta_V q_{AB}  =  \L_V q_{AB}-2 \alpha q_{AB} , \quad \delta_V C_{AB} =  \L_V C_{AB} - \alpha C_{AB} + \alpha u \dot{C}_{AB}  - 2 u (D_A D_B \alpha)^{\tf}
\end{array}
\end{equation}
Whence it naively appears as if a configuration $C^{(0)}$ which has zero news in say 
Bondi frame (where the associated $q_{AB}\ =\ \qo_{AB}$) goes to a new configuration $C^{(0)}+\delta_{V}C^{(0)}$ whose news is given by $\delta N_{AB}\ =\ 2 (D_{A}D_{B}\alpha)^{\tf}$.
However the above assertion is wrong as it relies upon the definition of news given by
\begin{displaymath}
N_{AB}(u,\hat{x})\ =\ -\partial_{u}C_{AB}(u,\hat{x}).
\end{displaymath}
This definition of news is only valid when the metric on $S^{2}$ is the unit metric $\qo_{AB}$. In a generic case  there is a slight technicality regarding the news tensor.\\ 

Given an arbitrary sphere metric $q_{AB}$ there exists a unique symmetric tensor $\rho_{AB}[q]$  which is implicitly defined via \cite{geroch}:
\be
\rho_{AB}q^{AB} = \R[q], \quad   D_{[A}\rho_{B]C}\ =\ 0 .
\ee
It can be split into a trace-free part and the trace part as 
\begin{equation}
\rho_{AB}\ =\ \rho^{(0)}_{AB} +\ \frac{1}{2}\R[q]q_{AB},
\end{equation}
and as shown in \cite{geroch} $\rho^{(0)}[\qo] =\ 0$.\\

The news tensor associated to a configuration $C_{AB}\ \in\ \Gamma^{q}$ is then defined as (see for instance Eq. (23) of \cite{newman})
\begin{equation}\label{newsdef}
N_{AB}(u,\hat{x})\ :=\ -\partial_{u}C_{AB}(u,\hat{x})\ - \rho^{(0)}_{AB}(\hat{x})  .
\end{equation}
We thus see that as $V\ \in\ \textrm{Lie}({\cal G})$ change $C_{AB}$ as well as $q_{AB}$, the corresponding change in news is given by
\begin{equation}
\delta_{V}N_{AB}(u,\hat{x})\ =\ -\partial_{u}\delta_{V}C_{AB}(u,\hat{x})\ -\ \delta_{V}\rho^{(0)}_{AB}(\xh).
\end{equation}
In section \ref{ssb1} we show that $\delta_{V}\rho^{(0)}_{AB}(\xh)$ is precisely such that a zero news configuration is mapped into a distinct zero news configuration. Whence the corresponding change in the news vanishes. That is \emph{any} element of $V\ \in\ \textrm{Lie}({\cal G})$ maps a configuration with zero news to a configuration with zero news because  the definition of the news before and after the action of $V$ refer to different frames.
Hence choosing a $q_{AB}$ (and working with  $\Gamma^{q}$) implies breaking the ${\cal G}$ symmetry spontaneously to $\textrm{BMS}$ and the subleading soft gravitons can be thought of as goldstone modes associated to this symmetry breaking as they map one family of vacua (associated to a given $q_{AB}$) to a distinct family of vacua associated to a different $q_{AB}$.

\subsection{Evaluating $\delta_{V}N_{AB}$}\label{ssb1}
From (\ref{deltaV}) and the definition of the news tensor (\ref{newsdef}) we have
\ba
\delta_V N_{AB} & =& - \partial_u \delta_V C_{AB}- \delta_V \rho^{(0)}_{AB} \\
& =& -\L_V \dot{C}_{AB}  - \alpha u \ddot{C}_{AB}  + 2  (D_A D_B \alpha)^{\tf} - \delta_V \rho^{(0)}_{AB}. \label{delVN2}
\ea
We now evaluate the last term in (\ref{delVN2}). From $\delta_V q_{AB}= \L_Vq_{AB}- 2 \alpha q_{AB}$  we have that $\delta_V \rho_{AB}$ is a sum of a Lie derivative term plus a scale transformation term. Since the behaviour of $\rho_{AB}$ under scale transformation is known \cite{geroch} the effect of the second term can be obtained explicitly. The total change is found to be:
\be
\delta_{V}\rho_{AB}= \L_V \rho_{AB} + 2 D_A D_B \alpha, \label{delrho}
\ee
which in turn implies,
\be
\delta_{V}\rho^{(0)}_{AB}= \L_V \rho^{(0)}_{AB} + 2 (D_A D_B \alpha)^\tf .\label{delrho0}
\ee
When substituting (\ref{delrho0}) in (\ref{delVN2}) the `soft' factors cancel out and one obtains:
\ba
\delta_V N_{AB} & =& - \L_V \dot{C}_{AB}    - \alpha u \ddot{C}_{AB} - \L_V \rho^{(0)}_{AB}  \\ 
&=& \L_V N_{AB} + \alpha u \dot{N}_{AB},
\ea
where in the second line we used the  definition of the news tensor (\ref{newsdef}) and the fact that $-\ddot{C}_{AB}= \dot{N}_{AB}$ since $\partial_u \rho^{(0)}_{AB}=0$.

Thus the news tensor transforms homogeneously. In particular if $N_{AB}=0$ then $\delta_V N_{AB}=0$.

\section{Conclusions} \label{sec5}
Analyzing the symmetry structure of the quantum gravity S-matrix is of paramount importance. It has been well known since the 60's that (at least at the semiclassical level) this symmetry group contains an infinite dimensional group known as the BMS group. The relationship of BMS symmetry to infrared issues in Quantum Gravity (for instance the existence of various superselection sectors) has been rigorously studied by Ashtekar  et. al. in the beautiful framework of Asymptotic Quantization \cite{aaprl,as,aajmp,aaoxf,aabook}. This relationship (of BMS group to infrared issues in quantum gravity in asymptotically flat spacetimes) got a new lease due to seminal work of Strominger et. al.  \cite{strominger0,supertranslation,virasoro}. One of the outcomes of this recent study is the universality concerning subleading corrections to soft graviton amplitudes, referred to in this paper as Cachazo-Strominger (CS) soft theorem \cite{cs}.\\
A natural question first posed in \cite{virasoro} was if the CS soft theorem could be understood as Ward identities associated to certain symmetries of the semi-classical S matrix. It was shown in \cite{virasoro} that the Ward identities associated to Virasoro symmetries contained in the so-called extended BMS group can be derived from CS soft theorem. However, the question of how to go in the reverse direction and derive the CS soft theorem from the Virasoro Ward Identities remained unanswered. In \cite{us} we argued in favor of a different possibility: If a different generalization of the BMS group (referred to unimaginatively as generalized BMS) ${\cal G}$ was a symmetry of the gravitational S matrix, then the Ward identities associated to $\diff(S^{2})$ contained in ${\cal G}$ were shown to be equivalent to CS soft theorem. Our argument however relied on an ad-hoc assumption that the charges associated to such symmetries had the same form as the charges associated to Virasoro symmetries, which (modulo certain IR issues) could be derived from first principles. The main block for computing such charges was lack of a suitable phase space on which ${\cal G}$ acted in a well-defined manner and whose corresponding charges were finite.\\
We have filled these gaps in the current paper. Starting from the covariant phase space associated to Einstein Hilbert action, we derive a phase space at null infinity which is coordinatized by the well known radiative degrees of freedom as well as the space of metrics on the conformal sphere. The symplectic structure on this phase space can be used to compute the charges associated to $\diff(S^{2})$ which is, rather remarkably well-defined on an appropriate subspace of the radiative phase space. Surprisingly \emph{these charges turn out to be exactly equal to the charges corresponding to the Virasoro symmetries computed in \cite{virasoro}.} This proves the key assumption that we  made in \cite{us} and hence completes the proof of the equivalence between Ward identities associated to the generators of $\diff(S^{2})$ and CS soft theorem.\\

One of the nice corollaries of our analysis is the representation of ${\cal G}$ on $\Gamma$. However the  symplectic structure arising from the Einstein Hilbert action is only well-defined in the stronger fall-offs subspace   $\Gamma_0 \subset \Gamma$  which unfortunately is not preserved under the action of $\G$.  We believe however that the inclusion of appropriate counter-terms to the action at $i^{0}, i^{\pm}$ could yield a well-defined symplectic structure on $\Gamma$. If this were to be the case, we could hope for the action of $\G$ to be symplectic on $\Gamma$. This would solve another issue which arose in \cite{us}, namely that the charges on radiative phase space which correspond to subleading soft factors do not close to form an algebra. We hope to come back to this point in the near future.  We  finally wish to emphasize that the physical phase space of the theory really is the radiative phase space (or an appropriate subspace thereof) and the bigger phase space $\Gamma$ is an ``auxiliary" arena which however is an indispensable tool to implement ${\cal G}$ in classical as well as quantum theory.\\

\noindent {\bf Acknowledgements}\\
We are indebted to Abhay Ashtekar for crucial discussions in the initial stages of this work and for his constant encouragement and interest. We thank Vyacheslav Lysov, Michael Reisenberger and Andrew Strominger  for their comments on the manuscript. MC is supported by Anii and Pedeciba. AL is supported by Ramanujan Fellowship of the Department of Science and Technology.

\appendix

\section{Closure of generalized BMS action} \label{appclosure}

The first relation (\ref{algebrafd}) is easily verified. We now show the second and third relations. \\

Let  $V_3:=[V_1,V_2]$. To shorten notation we will omit `$V$' labels and use only subscripts $1,2,3$. Thus the second equation in (\ref{algebrafd}) reads:
\be
[\delta_2,\delta_1]= \delta_3 , \label{123}
\ee
 and Equations (\ref{deltaV}) for $V_1$ become:
\ba
\delta_1 C_{AB}  & =&  \L_1 C_{AB} - \alpha_1 C_{AB} + \alpha_1 u \dot{C}_{AB}  - 2 u (D_A D_B \alpha_1)^{\tf} \\ \label{del1C}
\delta_1 q_{AB}  & = & \L_1 q_{AB}-2 \alpha_1 q_{AB} \label{del1q},
\ea
and similarly for $V_2$ and $V_3$.  For the computation it is important to keep in mind that the $\alpha$'s are in fact independent of the 2-metric $q_{AB}$ due to the condition $\sqrt{q}= \sqrt{\qo}$. This can be seen explicitly by defining $\alpha$ purely in terms of $\sqrt{q}$ according to:
\be
\L_V \sqrt{q} = 2 \alpha \sqrt{q}. \label{defalpha}
\ee
 We first verify (\ref{123}) along the $\delta q_{AB}$ direction:
\ba
[\delta_2,\delta_1] q_{AB} & = & \delta_2 (\L_1 q_{AB} - 2 \alpha_1 q_{AB} ) - 1 \leftrightarrow 2 \\
&=&  \L_1 \delta_2 q_{AB} - 2 \delta_2 \alpha_1 q_{AB} - 2 \alpha_1 \delta_2 q_{AB} ) -  1 \leftrightarrow 2 \\
&= &  \L_3 q_{AB} - 2( \L_1 \alpha_2 - \L_2 \alpha_1) q_{AB}  \\
&=& \delta_3 q_{AB}. \label{delta3q}
\ea
Here we used the fact that $\delta_2 \alpha_1=0$ since $\alpha_1$ is independent of $q_{AB}$ as mentioned above Eq. (\ref{defalpha}). In the last equality we used
\be
\alpha_3 = \L_1 \alpha_2 - \L_2 \alpha_1, \label{alpha3}
\ee
which directly  follows from the definition of $\alpha$ given in (\ref{defalpha}):
\be
2 \alpha_3 \sqrt{q} = \L_3 \sqrt{q} = (\L_1 \L_2 - \L_2 \L_1) \sqrt{q} = 2 \L_1 ( \alpha_2 \sqrt{q})  - 2  \L_2(\alpha_1 \sqrt{q}) =2 (\L_1 \alpha_2 - \L_2 \alpha_1) \sqrt{q}.
\ee
For the   $\delta C_{AB}$  direction,  one finds 
\begin{equation}
\begin{array}{lll}
[\delta_2, \delta_1] C_{AB} = \L_1 \delta_2 C_{AB} - \alpha_1 \delta_2 C_{AB} + \alpha_1 u \partial_u \delta_2 C_{AB} - 2 u \delta_2 (D_A D_B \alpha_1)^{\tf} - 1 \leftrightarrow 2  \\
=  \L_3 C_{AB} - \alpha_3 C_{AB} + u \alpha_3 \dot{C}_{AB} - 2u \L_1(D_A D_B \alpha_2)^\tf - 2 u \delta_2 (D_A D_B \alpha_1)^\tf-\\
\hspace*{4.5in}1 \leftrightarrow 2  . \label{del3C}
\end{array}
\end{equation}

where we used similar simplifications as when getting (\ref{delta3q}) above.  The `hard' term in (\ref{del3C}) corresponds to the hard term of $\delta_3 C_{AB}$. We now show that the `soft' term also matches. This amounts to show the equality:
\be
\delta_2 (D_A D_B \alpha_1)^\tf -  \L_2(D_A D_B \alpha_1)^\tf - 1 \leftrightarrow 2 = (D_A D_B \alpha_3)^\tf .\label{soft123}
\ee
The variation $\delta_2$ in (\ref{soft123}) only involve variations along $\delta q_{AB}$. It is convenient to write them as an explicit sum of `Lie derivative'  and `scale' terms:
\be
\delta_2 q_{AB} = \delta^L_2 q_{AB} + \delta^S_2 q_{AB} ;  \quad \delta^L_2 q_{AB} := \L_2 q_{AB}, \quad \delta^S_2 q_{AB}:= -2 \alpha_2 q_{AB}. \label{deltaLS}
\ee
In this way, the first term in (\ref{soft123}) takes the form:
\be
\delta_2(D_A D_B \alpha_1)^\tf = \delta^L_{21\; AB}+ \delta^S_{21\; AB}
\ee
where:
\ba
\delta^L_{21 \; AB}  & := & \delta^L_2 (D_A) \partial_B \alpha_1 -\frac{1}{2} \delta^L_2(\Delta) \alpha_1 q_{AB} - \frac{1}{2} \Delta \alpha_1 \delta^L_2 q_{AB} \label{delLAB} \\
\delta^S_{21 \; AB}  & := & \delta^S_2 (D_A) \partial_B \alpha_1 -\frac{1}{2} \delta^S_2(\Delta) \alpha_1 q_{AB} - \frac{1}{2} \Delta \alpha_1 \delta^S_2 q_{AB} \label{delSAB} 
\ea
For the $\delta^S$ term one obtains:
\be
\delta^S_{21 \; AB} = 2 D_{(A} \alpha_1 D_{B)} \alpha_2 - q_{AB} D_C \alpha _1 D^C \alpha_2. \label{delS21}
\ee
Since it is symmetric under $1 \leftrightarrow 2$ it does not contribute to the LHS of (\ref{soft123}).  For the $\delta^L$ term, we notice that from the definition of $\delta^L_2$ one has:
\be
\delta^L_2 D_A = [\L_2, D_A], \quad \delta^L_2 \Delta = [\L_2, \Delta], 
\ee
from which it follows that  (\ref{delLAB}) can be written as:
\be
\delta^L_{21\; AB} := \L_2 (D_A D_B \alpha_1)^\tf  -D_A \partial_B (\L_2 \alpha_1)+ \frac{1}{2} \Delta (\L_2 \alpha_1) q_{AB}. \label{delLAB2}
\ee
The first term in (\ref{delLAB2}) cancels the Lie derivative  term in (\ref{soft123}).  Including the $1 \leftrightarrow 2$ term one recovers Equation (\ref{soft123}) with $\alpha_3$ given in (\ref{alpha3}). This concludes the proof of Eq. (\ref{123}). \\

We finally show the last relation in (\ref{algebrafd}):
\be
[\delta_{V}, \delta_{f}] =-  \delta_{\L_V f- \alpha f} . 
\ee
Along $\delta q_{AB}$ direction this relation trivializes to $0=0$. Evaluating the commutator  along $\delta C_{AB}$ one finds
\begin{multline}
[\delta_f,\delta_V] C_{AB} = (\L_V f - \alpha f ) \dot{C}_{AB} \\+2 \alpha(D_A D_B f)^\tf +2 f (D_A D_B \alpha)^\tf + 2\delta_V(D_A D_B f)^\tf -2 \L_V(D_A D_B f)^\tf. \label{delfv}
\end{multline}
The `hard' term in (\ref{delfv}) matches the hard term of $\delta_{\L_V f - \alpha f}$. That the `soft' term (displayed in the second line) also matches can be shown along similar lines as for the soft term of $[\delta_{V_1}, \delta_{V_2}]$ computed above. Writing $\delta_V q_{AB} = \delta^L_V q_{AB} +\delta^S_V q_{AB}$ as in Eq. (\ref{deltaLS}) and using relations as those given in Eqns. (\ref{delS21}) and (\ref{delLAB2})  and finds that the last two terms in (\ref{delfv}) combine to
\be
2\delta_V(D_A D_B f)^\tf -2 \L_V(D_A D_B f)^\tf = -2 (D_A D_B \L_V f)^\tf +  4 D_{(A} \alpha D_{B)} f -2 q_{AB} D_C \alpha D^C f.  \label{df2}
\ee
The first term in the RHS of (\ref{df2}) is the soft factor of $\delta_{\L_V f}$. The remaining terms combine to give the soft factor of $\delta_{-\alpha f}$ due to the identity:
\be
(D_A D_B( \alpha f))^{\tf} = \alpha (D_A D_B f)^\tf + f (D_A D_B \alpha)^\tf + (2 D_{(A} \alpha D_{B)}f)^\tf .
\ee

\end{document}